\documentstyle[12pt]{article}
 \newcommand \be {\begin{equation}}
\newcommand \bea {\begin{eqnarray} \nonumber }
\newcommand \ee {\end{equation}}
\newcommand \eea {\end{eqnarray}}
 \newcommand \eps {\epsilon}
 \newcommand \s {\sigma}
\newcommand \de {\delta}
\newcommand \g {\gamma}
 \newcommand \al {\alpha}
 \newcommand \ran {\rangle}
 \newcommand \lan {\langle}

\begin{document}

\title{On the Replica Approach to Spin Glass \\
Talk given at Fifth International
Workshop on Disordered Systems, \\
 Andalo, February 1994}
\author{ Giorgio Parisi\\
Dipartimento di Fisica \\
Piazzale Aldo Moro, Roma 00187}
\maketitle

\begin{abstract}
In this talk I will review the approach to spin glasses based on the
spontaneously broken replica
symmetry. I will concentrate my attention mostly on more general
ideas, skipping technical details
and stressing the characteristic predictions of this approach. After
the
introduction of the replica method, the predicted structure of states
is investigated in details, paying a particular attention to the local
overlaps and to the structure of the clusters.
I will finally study the behaviour of the system near  the lower
critical dimension  and I will show that the technique of coupling real
replicas is able to give relevant information.
 \end{abstract}
\vfill
\vfill
\newpage

\section {Introduction}
The aim of this talk is to present some considerations on spin
glasses  in
the framework of the
broken replica approach.

There are many characteristic predictions of the replica approach
that
are in very good agreement
with the experiment. The most interesting prediction (in spectacular
agreement with the
experiments) is the following: the field cooled susceptibility in the
low
temperature glassy
region  is  independent from the temperature, on the contrary the
alternate susceptibility at fixed
frequency goes to zero  with the temperature.

However we would like to obtain more detailed predictions from the
theory and to compare them
with the very interesting experiments that have been done. Generally
speaking, if we consider  a
realistic finite dimensional spin glass, we  can firstly study it in
the
mean field
approximation and later on to compute the corrections to it.  While
in
many cases the fist step
has  been done, especially for the static, the second one has not yet
completed, so that many
results are  still conjectural.

In this note we will consider mainly models with the Hamiltonian of
this
form \cite{EA}:
\be
H=-\sum_{i,k} J_{i,k} \sigma_i \sigma_k - \sum_i h \sigma_i.
\ee

In the mean field approach \cite{mpv,parisibook2}, which should be
valid in the limit of infinite
number of neighbours  \footnote{More generally one neglects  the
existence of closed paths of finite
length on the lattice,
 which
implies  that $\sum_{k,m}J_{i,k}J_{k,m}J_{m,i}$ is much smaller than
$(\sum_{k}J_{i,k}^2)^{3/2}$.},
 the solution of a spin glass model is given by the broken replica
approach with the usual
ultrametric
hypothesis.

In this note we shall firstly recall the results obtained in the mean
field
approach and we shall
later discuss  how these results may extended to the short range
case.
In the second section we present a general introduction to the
replica method and
to the connection between the breaking of the replica symmetry and
the existence of many
equilibrium states. In the third section we study the structure of
the
equilibrium states in greater details. In section four we investigate
the
properties of the local overlaps, which give very crucial information
on the
properties of the system. In section five we introduce the concept of
cluster of
spins and we study its properties. In section six we present some
simple
predictions on the behaviour of the system near the lower critical
dimension,
which is  not very far from three. In section seven we show how
the
technique of coupling  real replicas may be used to characterise the
behaviour of
the
system.  Finally in the last section we present our conclusions.

\section {The General Setting}
In mean field approach there is an infinite number of stable states,
which are labelled by an index
 $\al$. They are characterised by a weight $w_\al= \exp (-\beta
f_\al)$,
$f_\al$ being (a part from
an addictive
constant) the total  free energy of the state $\al$. We will denote
by
$<>_\al$ the expectation in the
state
$\al$. The weights are normalised by the condition $\sum_\al w_\al=
1$.

It is usual to introduce their overlap and the energy overlap (in
zero
magnetic field:
\bea
q_{\al,\g}=\sum_i<\s_i>_\al<\s_i>_\al \ , \\
q^e_{\al,\g}=\sum_{i,k} J_{i,k}^2 <\s_i \s_k>_\al \ <\s_i \s_k>_\g \ .
\eea

For each given sample one can define the distribution probability of
the
overlaps
\bea
P(q) =\sum_{\al,\g} w_a w_g q_{\al,\g}, \\
P^e(q^e) =\sum_{\al.\g} w_a w_g q^e_{\al,\g}.
\eea
If replica symmetry is broken, the function $P(q)$ is non trivial.

At low temperatures the function $P(q)$ becomes concentrated at
$q=q_M$. For  $q \ne q_M$ the function $P(q)$ goes to zero at zero
temperature and it proportional to $T$. Indeed two different states
can  contribute to
the statistical sum
only if  they differ in energy of a quantity proportional to $T$; if
their
free energies  are
randomly distributed, this event happens with a probability which is
linear in
$T$.

There are two possible forms of the function $P(q)$ which correspond
to
quite   different situations. In the first case the function $P(q)$
is the
sum of  two delta
functions (i.e. $P(q)=   w_m\de(q-q_m)+w_M\de(q-q_M)$, while in the
second case the function
$P(q)$ contains a smooth  region  between  the two deltas. For the
moment we will restrict the
discussion to only this last  second case, in which all possible
values of
$q$ among $q_m$ and
$q_M$ are allowed.

The predictions I will present  are obtained in the mean field
approximation. The main physical
feature is the  existence of different equilibrium states, which are
not
related locally by any
symmetry  operation, which have the same free energy density
\footnote{These states have remarkable
properties, the most well known being ultrametricity
\cite{MV,MPSTV}.}.
  The existence of states with
free energy  difference of order 1 is a prediction which is valid in
the
mean field theory and could
disappear in more realistic short range model (although there is no
evidence for such an effect).

 Moreover no experiment or computer simulation on very large size
system
may  be sensitive to  these difference, so that it interesting to
find out
methods which test the
structure of replica  symmetry breaking, without to be sensitive to
these
details. This point
will be discussed in the next  sections.

\section{The Microstructure of the States}
In the mean field approach one finds that there is a strict
correlation
between the different kinds of
overlaps one can define. In particular in the case of the energy
overlap
one finds that
\be
q^e= A q^2 +B.
\ee
Consequently the probability distribution of the energy is given by
\be
P^e(q^e) = \left({q^e-B \over A}\right)^{-1/2}P\left({(q^e-B) \over
A}\right). \ee

In other words different equilibrium states do not differ only by the
reverse of large droplets of
spins \cite{FH}. If two states  differ only by the reverse of large
droplets of  spins,
the energy overlap would
be the same as that of a state with itself,  apart from the
contributions of
the interface among
different  clusters \cite{CPS}.

In the picture which emerge from replica symmetry
breaking
there  are clusters of  spins which are strongly coupled. These
clusters
have a surface which is
proportional to the volume so  that the change in energy overlap,
which
is proportional to the
surface, is comparable to the change in  overlap, which is
proportional
to the volume. Of course
we could call a fractal object a droplet, but this  would be a
linguistic
abuse.

In principle we can get information on the form
of the clusters from the analysis of the  correlations fucntions. The
situation  is rather complex
\footnote{See for example the description in the very interesting
paper
\cite{MV}}
and we shall only stress some features.

In order to define the correlation functions it may be useful
to consider two real replicas of the same system: the spins will be
denoted $\s$ and $\tau$
respectively. One can introduce the overlap  $q_i=\s_i\tau_i$. This
two
replicas system can be
characterised by the value of $q=1/V \sum_i q_i$. Different values of
$q$ in the interval $q_m -
q_M$ are allowed.
One can consider the connected correlation function  restricted on
configurations with a given overlap $q$:
\be
G_q(x)={1 \over V} \sum_i (<q_{i+x}q_i>_q- q^2 )\equiv {1 \over V}
\sum_i (<q_{i+x}>_q<q_i>_q- q^2).
\ee
The correlation function decreases in the mean field approximation in
dimensions $d$ as
\cite{DK1} \be
x^{-d+D},
\ee
where $D=3$, if $0 < q <q_M$, and  $D=4$, if $q=0$ (this is possible
only if $q_m=0$,
which  happens in zero magnetic field).

These results are supposed to be valid at all orders in the
perturbative expansion above  $6$  dimensions. The exponent $D$ may
change for less that 6 dimensions ($D$ at $q=0$ does  actually change
when $d<6$ \cite{DK2}).

Numerical simulations in $4$ dimensions \cite{PR,CPR} strongly
suggest the correctness of the broken
replica picture and no anomaly is observed.

\section{A Simple Model for the Clusters}
In order to interpret the previous results geometrically, we consider a
simple
model of spin glasses in
which  we find explicitly the existence of infinite number of
equilibrium
states \cite{ma,N}.
The model
is described by  the usual Hamiltonian, with an unusual distribution
of
the couplings:  if we  order the coupling in  decreasing strength we
must  have
that
\be
|J_a|>\sum_{b>a}|J_b|.
\ee

This can be realised by setting $J_{i,k}=(R_{i,k})^{g(L)}$, where
$R_{i,k}$ is a random number
uniformly distributed in the interval $[-1,1]$ and $g(L)$ is an
integer
function (which takes odd
values) which diverges  sufficiently fast when the size $L$ of the
box
goes to infinity.

We will consider this model at zero temperature. If we consider the
system in a box of size $L$
with  assigned boundary conditions,  there will be clusters of
rigidly
connected spins which are
controlled by the spins at the boundary. If we send the volume to
infinity, we find that the
ground state depends on  the boundary conditions.

The properties of the model may be found by mapping it on invasion
percolation. In dimensions $d>8$ one finds that these clusters of
rigidly  connected spinshave Hausdorff dimension 4 and the system  is
filled by an  infinite number of them. If we consider two systems with
random (and  different)  boundary  conditions, spins in different
clusters will be random oriented so  that the  global overlap will  be
zero.

The correlation function is given by the probability that two spins  at
distance $x$    belong to the same cluster. This probability  goes to
zero as $x^{- d+D_c}$, where $D_c$ is the  Hausdorff  dimension of the
cluster. If we stick to this  interpretation,  one finds that the
Hausdorff  dimension of the  clusters obtained from mean field theory
is  $4$, which is the same value as in this simple model.

In the real world things are more complicated. Clusters are not  really
independent, they  interact one with the othe. This
phenomenon may explain the  shift  of $D$ from $4$ to $3$  when we go
to non zero $q$. Lot of  theoretical work is needed to  present a full
understanding of  the properties of the clusters of strongly  connected
spins in the  framework of the broken replica  approach.

\section{Local Overlaps}

Generally speaking, the function $P(q)$ alone is a rather delicate
object  because its computation involves the  differences among  total
free energies  which are of order 1. We  could  consider situations in
which  the free energy differences are of order $L^\lambda$  with
$\lambda  <d$. In this a way the free energy differences diverge, but
the free energy densities are the same. In this situation  most of the
conclusions based on broken replica symmetry  would be valid, but the
function $P(q)$  would be trivial.

 It should be clear that a non trivial form of the function $P(q)$ at
zero  magnetic field is not  sufficient to rule out the droplet  model.
Indeed in the framework of  the  droplet model there could  be a few
compact  independent  droplets which can be flipped  independently.
Compact droplets  are characterised by the following  conditions:  the
radius $R$ goes to infinity the volume af a compact droplet goes as
$R^d$ and the surface as $R^{d-1+\omega}$   with $\omega<1$. If
$\omega>0$ the droplets have  a fractal surface. If $\omega=1$ they are
no more compact and in this  limit we recover the broken replica
approach.

In the droplet model the function $P(q)$ can be non trivial, but the
function $P^e(q^e)$ must go to a  delta  function acquiring a width of
order $L^{-\omega}$.

The differences among the broken  replica and droplet  model become
more sharp if consider the  probability distribution  $P_R(q_R|q)$ of
the local overlap  $q_R$ defined as
\be
q_R=\vert{1 \over R^d} \sum_k q_k\vert,
\ee
where the sum is done in a box of side $R$ centred around and
arbitrary point $i$\footnote{I am
grateful to J.P. Bouchaud for stressing this point to me.}. The
probability distribution may be
obtained  for a given sample by changing  the point $i$, keeping  the
constraint of total overlap $q$.

The quantity $q_R$ is invariant under a reversal of spins in a  cluster
that contains the whole  region of size $R$.  In the droplet model the
function $P_R(q_R|q)$,  should not depend on $q$ in  the infinite
volume limit: indeed when we reverse a large droplet,  the  overlap
$q_R$ changes only  for those points near the boundary of the droplet,
so that the  dependence  of $P_R(q_R|q)$ on $q$  should vanish as $({R
\over L})^{-1+\omega}$ when the size of system  $L$ goes to infinity.

\section{The Lower Critical Dimension}

In testing spin glass theories
in finite dimensions a crucial issue  is the  validity of the mean
field  picture. Now it seems that the mean field theory gives the
correct exponents in  dimensions ($d$)
greater than 6, while  some
corrections are expected in dimensions less than 6.  Unfortunately  the
computation of the  corrections to mean field theory is rather complex
and the one loop  contribution has not been fully  computed so that no
precise conclusions are available.

The most critical issue is if the behaviour of  the propagator at $q\ne
0$ (i.e. $1/k^3$ in momentum space) remains  unchanged. This behaviour
arises as a consequence of the breaking of replica symmetry in a
continuos way. The corresponding  Ward identities has not been
carefully spelled out, so that it is not  evident if such a relation is
renormalized (the $1/k^4$ behaviour at $q=0$ apparently changes in
less that $6$ dimensions).

If this behaviour would be not renormalized, it would be an easy  guess
to suggest that the lower critical  dimension of the system is $3$ in
the same way as the lower  critical dimension for breaking of
continuos symmetries is 2. A direct computation of the  free  energy of
an interface among states at  different values of $q$ gives rather
different  results,  i.e. a lower critical dimension of $2.5$,  which
is by {\it accident} the same obtained in the  framework of the Migdal
Kadanoff approximation \cite{FPV2}.

The discrepancy among the two different results has an unclear  origin,
which we hope to be able  to understand in the next future, when the
evaluation of the loop  corrections to various different  quantities
will be available. In any case  3 it is not very far from  being  the
lower critical dimension so that it is reasonable to investigate what
should happen  near  the critical dimension and to compare with the
experimental (or numerical results).

The renormalization group scheme for the behaviour at the lower
critical dimension has not yet developed. Let us try do some educated
guesses. We  suppose that for  a  system  with continuos distribution
of the couplings $J$  \footnote{Anomalous behaviour may  be present if
the couplings are discrete when $T \to 0$, due  to the degeneracy of
the ground state},  there is only one
 relevant quantity that describe the behaviour of the system at scale
$L$  and it
 satisfies a renormalization group equation
\be
{dT \over dt}= \beta(T(t)).
\ee
The quantity $T(t)$ can be identified with the effective temperature
at
scale $L=\ln(t)$ and the
function $\beta(T)$ should  behave as a power of $T$ as small $T$.

 One readily finds that if
\be
\beta(T) \propto T^\lambda,
\ee
the correlation function diverges as
\be
\xi \propto \exp ( a/T^{(\lambda-1)}).
\ee
In the Migdal Kadanoff one finds that approximation $\lambda=3$.

In the same way we expect that the field $q$ has an effective
dimension
($\psi(T)$) which is
proportional to $T$, and therefore the value of $q$, observed on a
scale
of size $L$, should
decrease (roughly speaking) as
\be
q(L) \propto L^{-\psi(T(\ln(L))}.
\ee

 In a given range of scales, which may be quite wide at low
temperatures if $\lambda$ is large,  $T(t)$ may be approximated with a
constant. If this happens in the  low  temperature region, where the
correlation length $\xi$ is much greater that the side of  the  system,
the correlation function should decrease with temperature  dependent
power law. The range of validity of these approximate power law
depends on the  form of the  function $\beta(T)$. In the two dimension
ferromagnetic $x-y$ model the $\beta$  function is  identically zero so
that a power behaviour is exact.

Under the approximation of a constant $T(t)$, the
function $P(q)$ in  a  box of size $L$ is a  function of
$q L^{-\psi(T)}$, the space correlation function of two overlaps ($\lan
q(x) q(0) \ran$) decays as $x^{-2  \psi(T)}$  and the time  correlation
function  of two spins ($\lan \s(0,t) \s(0,0) \ran$
decays as $t^{\psi(T)/z(T)}$, where $z(T)$ is a  temperature dependent
dynamic exponent. Some numerical indications of such a behaviour in
three dimensions are  already been obtained and further work is needed
to distinguish the  case  of a system at the lower  critical dimension
or slightly above it \cite{MPR1,LMPR}.

These predictions can be tested relatively easily within a dynamic
study.  Let us consider the  dynamics of two replicas starting by two
different boundary  conditions.   The two replicas will have  zero
overlap for times not  very large compared with the volume. In  this
way the dynamics automatically  selects the state $q=0$. In the same
way, by observing  the  same single system at two large times  $t_1$
and $t_2$, such that ${|t_1-t_2| \over t1+t_2|}  <<1$,  the two
configurations at different  times should behaviour as two generic
configurations a  $q=q_M$ \cite{CKP}.

Large scale simulations have been recently done on a three
dimensional
system and high precision
data have been obtained in the high temperature region. One finds out
that the data are compatible
both with a  phase transition at a non zero temperature and with a no
transition scenario , altough this last scenario seems more consistent.
\cite{MPR2}.

There are some disturbing feature in the framework of the transition
scenario. Indeed one could
look (in a two replica system) to the value of the Binder cumulant
\be
g= {K-3 \over 2}; \ \ \ \ K= {<q^4> \over <q^2>^2},
\ee
where the average is also done over the different realisation of the
systems.

The Binder cumulant can be computed for different sizes and the value
where the different curves
intersect, characterises the critical temperature. The value of $g$
at the
critical point should be
universal. As far as the curves for the Binder cumulant collapse at a
given temperature (as
predicted from replica symmetry breaking), it is difficult to
establish
the existence of a point in
which the curves merges, because a small shift is enough to avoid the
intersection.

There is good numerical evidence that the value of $g$ at the
apparent
intersection point is
strongly dependent on the  choice of the probability distribution for
the
couplings $J$
\cite{MPR2}.

This dependence of $g$ on the probability distribution of the
couplings
strongly suggest that $3$ is
the lower critical dimension or it is very near to it and the real
critical
temperature, if any, is
far  from the apparent point where different cumulants merge.

It should also be said that simulations (and experiments) at
temperature
smaller that the apparent
critical temperature are well compatible with the possibility that
$D=3$
is the lower critical
dimension.

Very large scale simulations with a dedicated computer \cite{OGI} have
shown  a power  law decay of the
correlation of the same spin at different times: $C(t)\propto t^{-
\delta}$. Real experimental data
show a behaviour similar behaviour for the remanent magnetization as
function of the waiting
time. They are compatible with the law $M(t,t_W)=t^{-\delta}
f(t/t_w)$. The presence of a non zero
value of $\delta$ can be considered as an effect of having an
anomalous
dimension for the $q$ field.

Other simulations for the static properties done in the low
temperature
region seem to be well
compatible with the scenario where no transition is present and that
$q$
has a temperature dependent
dimension.

\section{Coupled Replicas}
Some of the previously discussed points become clear if we consider
the following Hamiltonian \cite{CPS,FPV1}:
\be
H=-\sum_{i,k} J_{i,k} (\sigma_i \sigma_k +\tau_i \tau_k) - h \sum_i
(\s_i+\tau_i)
-\eps \sum_i  \sigma_i \tau_i.
\ee
The term proportional to $\eps$ has the tendency to align the two
replicas in the same direction
for positive $\eps$ and in the opposite direction for negative
$\eps$.
Therefore we expect that
\bea
\lim_{\eps \to 0^+} q(\eps) = q_M \\
\lim_{\eps \to 0^-} q(\eps) = q_m
\eea

An unexpected result  of a detailed computation \cite{FPV1} show that in
the  replica
approach
\bea
q(\eps) = q_M +A \eps^{1/2} \ \  for \ \ \eps>0 \\
q(\eps) = q_m - B \eps^{1/2} \ \  fo r\ \  \eps>0.
\eea
In other words both the spin glass susceptibility and the $q - q$
correlation length diverge when
$\eps \to 0$.

This prediction of the replica approach is very interesting because
it can
be well studied in
computer simulations as far as the function $q(\eps)$ can be computed
with a reasonable amount of
time, at leas for not infinitesimal $\eps$. The correlation function for
$q$ is finite for non zero $\eps$.  In
other words the new term
forces the two systems to be in the same state and also if the state
is not
the real equilibrium one its
properties should no be crucial dependent on this fact.

The divergence of the correlation length when $\eps \to 0$ implies
the
possibility of changing
larger and larger clusters of spins and the consequent divergence of
the
correlation time.

The naive replica predictions seems to be approximately well
satisfied in
4 dimensions \cite{PR,CPR},
while they  are far from being correct in the three dimensions
\cite{MPR2}. In this case at zero
magnetic field the  preliminary data are compatible with a law of the
type
\be
q=\eps^{\lambda(T)},
\ee
where the exponent $\lambda(T)$ is related to the exponent $\psi(T)$
of section (5) by
\be
 (\lambda+1)  \psi(T)= 1.
\ee

\section{Conclusions}

In the recent years there have been both
theoretically
and experimentally progresses in the study of spin glasses. For short
range models the theoretical situation is still not completely clear
because we do not  known how to compute
systematically the corrections to the mean field theory (or better we
know how to do, but the
computations are too complex).

Experimentally (from numerical simulations) it seems that the mean
field picture is in very nice
agreement with the results in four dimensions and in three dimensions
for not too large size.

 An
open possibility is that the critical dimension is near to 3. In this
case
there would be a wide
temperature range where the correlation length is very large, but
finite.
The
predictions from replica symmetry breaking on the global order
parameters should apply to systems of
size smaller then the correlation length.  All the data I know are
consistent with this possibility,
which should be better investigated numerically and theoretically. I
hope that a strong progress will
be done in the next future on this particular point.

 \section{Acknowledgements}
It is pleasure for me to thank for very useful discussions and
fruitful
collaboration on these
problems L. Cugliandolo, S. Franz, J. Kurchan, E. Marinari, F. Ritort
and M. Virasoro.

 \end{document}